\documentclass[12pt]{iopart}
\usepackage{epsf}
\usepackage{epsfig}
\usepackage{graphicx}
\usepackage{dcolumn}
\usepackage{bm}

\newcommand{\la}{\langle}
\newcommand{\ra}{\rangle}
\newcommand{\ua}{\uparrow}
\newcommand{\da}{\downarrow}

\begin{document}
\title{Weak ferromagnetism and spiral spin structures in  honeycomb Hubbard planes}
\author{M A  N Ara\'ujo$^{1,2}$ and N M R Peres$^{2,3}$}

\address{$^1$Departamento de F\'{\i}sica, 
Universidade de \'Evora, P-7000-671, \'Evora, Portugal}
\address{$^2$Center of Physics, 
Universidade do Minho, P-4710-057, Braga, Portugal}
\address{$^3$Department of Physics, Boston University, 590 
Commonwealth Avenue, Boston, MA 02215,USA}
\ead{mana@uevora.pt}

\begin{abstract} 
Within the Hartree Fock- RPA analysis, 
we derive the spin wave spectrum for the weak ferromagnetic
phase of the Hubbard model on the honeycomb lattice.
Assuming a uniform magnetization,
the polar (optical) and acoustic branches of the spin wave excitations
 are determined. The bipartite lattice geometry produces a $\bm q$-dependent
 phase difference between the spin wave amplitudes 
on the two sub-lattices.
We also  find an instability of the uniform weakly 
magnetized configuration to a weak antiferromagnetic spiraling 
spin structure, in the lattice plane, with  wave vector ${\bm Q}$ 
 along  the $\Gamma-K$ direction, for electron densities $n>0.6$.  
We discuss the effect of diagonal disorder on both the creation
of electron bound states,  enhancement of the density of states,
and the possible relevance of these effects  to disorder induced
ferromagnetism, as observed in proton irradiated graphite.

\end{abstract}
\submitto{\JPCM}
\maketitle

\section{Introduction}
Recent interest in strongly correlated systems in non-square lattices,
such as the triangular, honeycomb and kagom\'e lattices, 
is justified by  the possible realization of 
exotic metallic \cite{anderson,kopelevich1},
magnetic \cite{matthew,esquinazi} and superconducting  states \cite{tanaka}
both in inorganic and organic materials.
From the organic side, graphite and related
carbon allotropes are physical systems where growing evidence
for exotic types of ground states is being accumulated during the last
few years.    In graphite, for example, experimental research 
 put forward evidence for unusual metallic and
magnetic properties \cite{kopelevich1,kopelevich2,makarova,esquinazi}.
In particular, ferromagnetism 
has been observed at high temperature in graphite \cite{kopelevich2}
which may not be due to magnetic impurities. Also, the 
observation  \cite{kopelevich1}
of magnetic order induced  by proton irradiation
challenges the theoretical description. Graphite
is not alone on the ferromagnetic-order-by-disorder scenario, with 
the inorganic CaRuO$_3$ material also exhibiting disorder-induced 
ferromagnetism \cite{he}. Recent experimental work
\cite{novoselov,zhang,berger} has produced atomic thin graphite 
planes where the exciting physics of 2D Dirac 
fermions may be directly observable.
Motivated by these experimental 
studies and because
the microscopic origin of ferromagnetism in these
compounds is far from being  understood, 
we decided to study the magnetic
properties of a doped  Hubbard model on an honeycomb lattice --
a single graphite plane.
To the  best of our knowledge, ferromagnetic spin waves 
in the honeycomb lattice (as an  itinerant electron system) have  not 
been studied in the past.
The fact that the honeycomb lattice
is a Bravais lattice with a basis immediately presents us the 
possibility of observing both polar and acoustic spin waves
\cite{mattis,yamada}.
We focus our research on the stability of the 
weak homogeneous ferromagnetic phase
found in \cite{prb04}.
 The paper is organized as follows: the model is 
introduced
in section \ref{modelsec} and the energy spectrum for 
the weak homogeneous ferromagnetic system is derived;
Section \ref{spinwvsec} is devoted to the description 
of spin waves in the 
weak homogeneous ferromagnetic phase;  
the possibility of spiral  spin states is investigated
in Section \ref{sprls}, 
where a spiral arrangement is 
found with lower energy than 
that of a uniform magnetization;
 Section \ref{disorder} gives a 
discussion on the possibility
of formation of electronic bound states, due to  impurities, and of
the possible relevance of disorder
to experiments on proton irradiated graphite.

\section{Model Hamiltonian}
\label{modelsec}

\begin{figure}[ht]
\begin{center}
\epsfxsize=8cm
\epsfbox{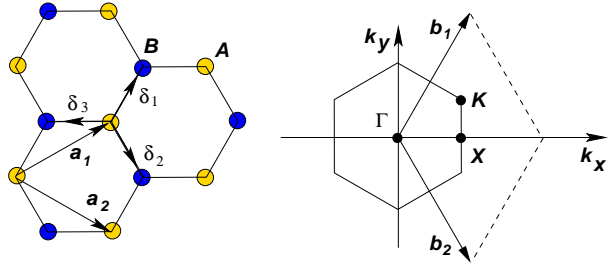}
\end{center}
\caption{ Primitive vectors ($\bm a_1$ and $\bm a_2$)
for the honeycomb lattice. The  vectors $\bm \delta_1$, $\bm \delta_2$ and $\bm \delta_3$ connect the A site to its 
three  neighbouring B sites. 
The hexagonal first  Brillouin zone corresponds to the reciprocal lattice vectors, $\bm b_1$ and $\bm b_2$.}
\label{honey}
\end{figure}

The Hubbard model is defined as 
\begin{equation}
H = -\sum_{i,j,\sigma} (t_{i,j}+\mu\delta_{ij}) 
c^{\dag}_{i,\sigma}  c_{j,\sigma}
+U\sum_i
  c^{\dag}_{i,\ua}c_{i,\ua}  c^{\dag}_{i,\da}  c_{i,\da}\,,
\label{hubbard}
\end{equation}
where $c^{\dag}_{i,\sigma}$ $(c_{i,\sigma})$ 
represents a creation (destruction)
electron operator with spin $\sigma$ at site $i$, $t_{i,j}$ is
the hopping integral between two sites $i$ and $j$, $U$ is the on-site
Coulomb repulsion, and $\mu$ is the chemical potential.
In the honeycomb lattice we identify two sub-lattices, $A$ and $B$ 
(see figure \ref{honey}), where the 
 primitive vectors of the underlying triangular lattice are denoted by 
$\bm a_1$ and  $\bm a_2$. For later use we  also define the vector $\bm a_3 \equiv (\bm a_1-\bm a_2)$.
 The reciprocal lattice vectors are $\bm b_1$ and $\bm b_2$ and define a hexagonal shaped first Brillouin Zone.  
Also shown in figure \ref{honey} are the vectors connecting any $A$ atom to its nearest
 neighbours,  denoted as $\bm \delta_1$, $\bm \delta_2$ and $\bm \delta_3$.  
 The electrons leaving on each sub-lattice will be 
denoted by field operators $a$ and $b$, respectively. 
The Fourier transformation between real and  momentum spaces is given by: 
\begin{equation}
  a^\dag_{i,\sigma}=\frac {1}{\sqrt {N_A}}\sum_{\bm k}e^{i\bm k\cdot \bm R_i}
  a^\dag_{\bm k,\sigma}\,,\hspace{.2cm}
  b^\dag_{i,\sigma}=\frac {1}{\sqrt {N_B}}\sum_{\bm k}e^{i\bm k\cdot \bm R_i}
  b^\dag_{\bm k,\sigma}\,,
\end{equation}
and we take $N_A=N_B=N$ as the number of unit cells. In the 
calculations below,  we shall consider first
 and second neighbour hopping integrals, $t$ and $t'$, respectively. 
The Hubbard model then takes the form:
\begin{eqnarray}
H&=&\sum_{\bm k,\sigma} (D(\bm k)-\mu) (a^\dag_{\bm k,\sigma}a_{\bm k,\sigma}+
b^\dag_{\bm k,\sigma}b_{\bm k,\sigma})
\nonumber\\
&+&\sum_{\bm k,\sigma}[\phi(\bm k)a^\dag_{\bm k,\sigma}b_{\bm k,\sigma}+
\phi^\ast(\bm k) b^\dag_{\bm k,\sigma}a_{\bm k,\sigma}] + H_U\,,
\label{hamilt}
\end{eqnarray}
with 
\begin{eqnarray}
D(\bm k)&=&-2t' \sum_{i=1}^3 \cos(\bm a_i\cdot {\bf k})\,, \nonumber\\
\phi(\bm k)&=&-t\sum_{i=1}^3 e^{i\bm k\cdot \bm\delta_i}\,.
\end{eqnarray}
In the  ferromagnetic ground state, the average occupancy of lattices sites
is given by
\begin{equation}
\la a^\dag_{i,\sigma}a_{i,\sigma}\ra = \frac n 2 + \sigma \frac m 2\,, 
\hspace{0.5cm}
\la b^\dag_{i,\sigma}b_{i,\sigma}\ra = \frac n 2 + \sigma \frac m 2\,, 
\label{broke}
\end{equation}
 with  the spin index $\sigma=\pm 1$. 
This  may  also be generalized to describe 
antiferromagnetic ordering if we replace  $m$ with $-m$ in one of the equations
(\ref{broke}) \cite{prb04}.  An Hartree-Fock treatment 
of  the Hubbard term, $H_U$,
taking into account equation (\ref{broke}), yields a set of 
quasi-particle bands given by
\begin{equation}
E_{\sigma}^{\alpha}(\bm k)=D(\bm k)+\frac U 2 (n-\sigma m)+\alpha 
\vert\phi_{\bm k}\vert\,,
\end{equation}
where $\alpha=\pm$ is a  band index. In the ferromagnetic
phase the single particle Green's functions can be written, in
 momentum space, as:
\begin{eqnarray}
{\cal G}_\sigma^{aa}(i\omega_n,{\bm k})&=& \sum_{j=\pm}
\frac{1/2}{i\omega_n - E^j_\sigma(\bm k)}
\label{gaa}\\
{\cal G}_\sigma^{ab}(i\omega_n,{\bm k})&=& \sum_{j=\pm}
\frac{je^{i\delta(\bm k)}/2}
{i\omega_n - E^j_\sigma(\bm k)}\label{gab}\\
{\cal G}_\sigma^{ba}(i\omega_n,\bm k)&=& \sum_{j=\pm}
\frac{je^{-i\delta(\bm k)}/2}
{i\omega_n - E^j_\sigma(\bm k)}\label{gba}\\
{\cal G}_\sigma^{bb}(i\omega_n,{\bm k})
&=& {\cal G}_\sigma^{aa}(\omega_n,\bm k)\,,
\label{gbb}
\end{eqnarray}
where we have defined $e^{i\delta(\bm k)}=\phi(\bm k)/\vert\phi(\bm k)\vert$.
The Hartree-Fock magnetization is given by 
$
m= (1/2N)\sum_{\bm k,\alpha,\sigma}\sigma f[E_\sigma^\alpha(\bm k)]$.

Since we are mainly concerned with the ferromagnetic phase, the calculations
in sections \ref {spinwvsec} and \ref{sprls}
 are performed for an electronic density smaller than
half filling. Therefore, electrons at the Fermi level will not  be treated  
as massless  Dirac fermions. Such treatment    is usually appropriate
for electrons in graphite planes at half filling
(or close to half filling) \cite{guinea}. 

\section{Magnetic collective excitations}
 \label{spinwvsec}

We obtain the   magnetic collective excitations  from  the poles
of  the transverse spin susceptibility calculated in the RPA approximation.
 Because there are two sub-lattices, the susceptibility is 
actually a second order tensor given by the expression
\begin{equation}
\chi^{i,j}_{+-}(\bm q,i\omega_{n})=
\int_0^{1/T} d\tau e^{i\omega_{n}\tau}\la T_{\tau} 
  S^{+}_i(\bm q,\tau)  S^{-}_j(-\bm q,0) \ra
\end{equation}
where $i,j=a,b$ are sub-lattice labels
and $S^{+}_i(\bm q),\  S^{-}_j(\bm q)$ are  the
spin-raising and lowering operators for each sub-lattice. 
The RPA expansion  gives a  Dyson equation for the transverse 
spin susceptibility, 
which can by written, in matrix form, as 
\begin{eqnarray}
 \bm\chi (\bm q,i\omega_n)
= \Big[ \bm 1 -  \frac{U}{N}\bm\chi^0(\bm q,i\omega_n) \Big]^{-1} 
\bm\chi^0(\bm q,i\omega_n)
\end{eqnarray}
where $\bm 1$ denotes the $2\times 2$ identity matrix. 
The poles of the susceptibility tensor, corresponding to the
magnetic excitations, are then obtained from the condition:
\begin{equation}
{\rm det} \Big[ \bm 1 
-  \frac{U}{N}\bm\chi^0(\bm q,\omega+i0^+) \Big] = 0.
\label{det}
\end{equation}
Below the particle-hole continuum of excitations, the 
spectral (delta-function contributions) part in
$\chi^{(0)ij}_{+-}(\bm q,\omega  +i0^+)$ vanish and 
there is the additional relation 
$\chi_{+-}^{(0)ba}=(\chi_{+-}^{(0)ab})^\ast$.
The zero order susceptibility tensor, 
$\bm \chi_{+-}^{(0)}(\bm q,\omega)$, can be written as
\begin{equation}
\bm \chi_{+-}^{(0)}(\bm q,i\omega_n)=-\frac{1}{4}
\sum_{\bm k;\alpha_1,\alpha_2=\pm}
\frac {f_\ua^{\alpha_1}(\bm k)-f_\da^{\alpha_2}
(\bm k-\bm q)}{i\omega_n+E^{\alpha_1}_\ua(\bm k)-
E^{\alpha_2}_\da (\bm k-\bm q)} {\cal A}(\bm k, \bm q)\label{chi}
\end{equation}
where the matrix $ {\cal A}(\bm k, \bm q)$ contains the coherence factors: 
\begin{equation}
{\cal A}(\bm k, \bm q)=\left(\begin{array}{cc}
1 & \beta(\bm k, \bm q) \\
\beta^*(\bm k, \bm q) & 1\end{array}\right)\,,
\end{equation}
 $\beta(\bm k, \bm q) = \exp[-i\delta(\bm k)+i\delta(\bm k-\bm q)]$,
with $f_\da^{\alpha_2}(\bm k)$ representing the Fermi function with argument
$E_\da^{\alpha_2}(\bm k)$, and equivalent representations hold for the
other cases.

\begin{figure}[htf]
\begin{center}
\epsfxsize=8cm
\epsfbox{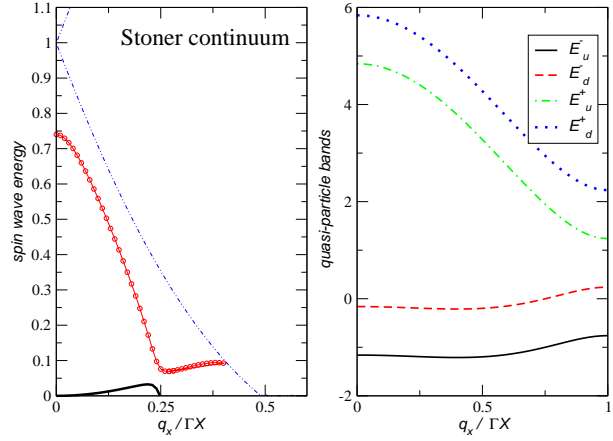}
\end{center}
\caption{
Left panel: Spin waves (line with circles for $\omega_{opt}$
and continuous line for $\omega_{ac}$) and Stoner continuum border
(thin dashed-dotted line) along the
$\Gamma-X$ direction of the Brillouin zone. Right: 
band energies along the
$\Gamma-X$ direction  (the band energies are measured 
relatively to the chemical potential);
The parameters are: $U=4$, $t=1$, $t'=-0.2$, $n=0.75$,
and the magnetization
and chemical potential are $m=0.25$ and $\mu=0.36$.
The energy through all this paper is in units of $t$.
(Subscripts $u$ and $d$ in the  right panel
denote spin projections $\ua$ and $\da$, respectively.)}
\label{sw1}
\end{figure}

\begin{figure}[htf]
\begin{center}
\epsfxsize=8cm
\epsfbox{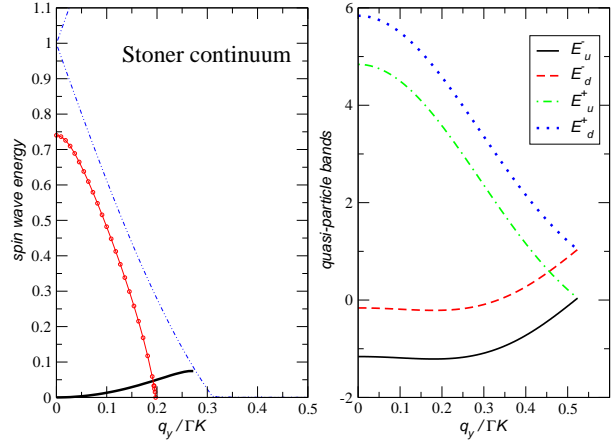}
\end{center}
\caption{ 
Left panel: 
Spin waves (line with circles for $\omega_{opt}$
and continuous line for $\omega_{ac}$) and Stoner continuum border
(thin dashed-dotted line) along the
$\Gamma-K$ direction of the Brillouin zone. Right panel: 
quasi-particle band structure along the
$\Gamma-K$ direction (band energies measured 
relative to chemical potential);  
The parameters are the same as in figure \ref{sw1}.
(Subscripts $u$ and $d$ in the   right panel
denote spin projections  $\ua$ and $\da$, respectively.)}
\label{sw2}
\end{figure}

In addition to the collective ferromagnetic spin waves there are also 
 single particle flip-spin excitations which define the so
called Stoner continuum.
In our case we may have up to four regions in the energy-momentum plane
associated with the latter type of excitations. Their spectra are given by
\begin{equation}
\Delta^{\alpha_1,\alpha_2}(\bm q)=E^{\alpha_1}_\da(\bm k-\bm q)-
E^{\alpha_2}_\ua(\bm k)\,.
\label{stoner}
\end{equation}
Depending on the position of the Fermi level, one or two
of the regions defined by equation (\ref{stoner}) may not occur
because the bands may be empty.
Because we have the lower ($E^{-}_\sigma$) bands partially filled, 
there always are 
two Stoner regions given by $\Delta^{-,-}(\bm q)$ and $\Delta^{+,-}(\bm q)$. 
 We consider only this case below (for the other cases the results are
qualitatively the same). In figures \ref{sw1} and \ref{sw2} the left
panel shows the the Stoner continuum defined by  $\Delta^{-,-}(\bm q)$
as the area enclosed by the dashed-dotted line  starting at 
$\Delta^{-,-}(0)=Um$. The model parameters have been chosen  
so that the system is definitely not  antiferromagnetic.
In figures \ref{sw1} and \ref{sw2} we  plot the solutions to
equation (\ref{det}). It is clear that the system has two different 
types of collective magnetic excitations, namely 
the usual acoustic mode
$\omega_{ac}(\bm k)$ and the polar or optical mode $\omega_{opt}(\bm k)$,
associated with the existence of two atoms per unit cell \cite{mattis,yamada}.
It is quite interesting that the behaviour of both branches of excitations
does not follow the same trend along 
different directions of the Brillouin zone:
 along the $\Gamma-X$ direction (see figure \ref{sw1}), for example, and for
 $\vert\bm q\vert> 0.25\Gamma X$ 
only the optical branch remains. On the other hand, in the 
$\Gamma-K$
direction  (see figure \ref{sw2}) it is the optical branch that vanishes
 at  $\vert\bm q\vert\approx 0.2\Gamma K$ while the acoustic mode survives.
The vanishing frequency of the acoustic (or optical) modes at finite momentum 
is associated with an instability of the homogeneous weak ferromagnetic phase 
toward a state exhibiting possibly weak ferromagnetic order 
in the $z$ direction and spiral order in the $xy$ plane, 
which will be analyzed in section \ref{sprls} below.

The eigenvector of the matrix $\bm\chi^{(0)}(\bm q$, $\omega(\bm q))$ that is
associated with the  eigenvalue $N/U$  gives the 
spin wave amplitudes over the $A$ and  $B$ sub-lattices. 
We note that equation (\ref{det}) is equivalent to the eigenproblem
\begin{equation}
\frac 1 N \bm\chi^0(\bm q,\omega+i0^+) \left[ \begin{array}{c}
 \la S_A^+\ra  \\ \la S_B^+\ra  
\end{array}\right]
= \frac 1 U \left[\begin{array}{c}
 \la S_A^+\ra  \\ \la S_B^+\ra 
\end{array} \right]\,,
\label{spinor}
\end{equation}
where $\la S^+_A\ra ,\la S^+_B\ra $ denote the spin wave amplitudes over the two
sub-lattices \cite{tesa}.
In our case,
$\chi^{(0),aa}=\chi^{(0),bb}$ and $\chi^{(0)ab}=(\chi^{(0)ba})^*$. 
The equations for the eigenvalue $\lambda$ and the corresponding eigenvector
 are:
\begin{eqnarray}
\lambda &=& \chi^{(0),aa} \pm \vert \chi^{(0)ab} \vert \\
\la S^+_A\ra &=& \pm \frac{ \chi^{(0)ab} }{ \vert \chi^{(0)ab} \vert} \la S^+_B\ra \,,
\label{angle}
\end{eqnarray}
where $\lambda=N/U$ is the relevant eigenvalue,  as can be seen from equation (\ref{spinor}).
 Equation (\ref{angle}) shows that 
the  spin wave amplitudes  are related by a phase factor.  
Therefore, the phase of the complex matrix element $ \chi^{(0)ab}$ determines
the angle between the transverse spin components  $\la S^+_A\ra $ and  $\la S^+_B\ra $.
The optical mode in the $\Gamma-X$
  direction (shown in figure \ref{sw1}) starts off with $\la S^+_A\ra =-\la S^+_B\ra $
for small $\bm q$, as expected of an optical mode, but
the angle between  $\la S^+_A\ra $ and  $\la S^+_B\ra $
 monotonically  decreases from $\pi$,
upon increasing  wave vector, and equals $\pi/2$ when $\omega_{opt}$  attains 
its minimum. At that point,  $\chi^{(0)ab}$ is pure imaginary.
\begin{figure}[htf]
\begin{center}
\epsfxsize=8cm
\epsfbox{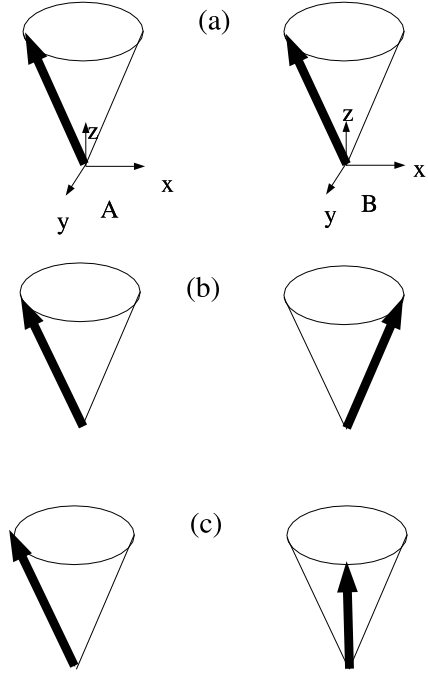}
\end{center}
\caption{
Relative position of the precessing spins on neighbouring $A$ and $B$ sites
for a spin wave in the $\Gamma-X$ direction: 
$(a)$ acoustic mode with small $\bm q$; $(b)$ optical mode with small $\bm q$; 
$(c)$ minimum (maximum) frequency of the optical (acoustic) mode.}
\label{cones}
\end{figure}
On the other hand, in the acoustic mode  the angle increases from zero to
$\pi/2$, when $\omega_{ac}$ is maximum  (the angle between the spins is illustrated
 in figure \ref{cones}).
 As the wave vector further increases, the acoustic mode frequency 
rapidly  decreases and vanishes at   $\bm q \approx 0.25 \bm{ \Gamma X}$.
 
Considering now the spin waves in the 
$\Gamma-K$ direction, the angle between the precessing spins  is
always zero or $\pi$ in  the acoustic and optical modes, respectively.
The optical mode frequency vanishes shortly after the interception with
 the acoustic branch and only the latter survives for increasing  $\bm q$. 

The phase difference between  spin wave amplitudes just described for
 the $\Gamma-X$ direction
 is a manifestation of the  complex coherence factors  appearing in the single 
particle  Green functions (\ref{gab})  and (\ref{gba}), 
resulting from  the honeycomb lattice geometry.

\section{Spiral spin states}\label{sprls}

The disappearance of the acoustic mode, at 
wave vector ${\bm q}\approx 0.25  \bm{\Gamma X}$,
and  of the optical mode, at 
wave vector ${\bm q}\approx 0.20  \bm{\Gamma K}$,
suggests  an instability to a spiral spin state \cite{Auerbach}. 
In such a state,
the spiral spin configuration is characterized by non-zero 
transverse magnetization
 at site $j$, $\la S^+_j\ra =\la S^+\ra e^{i{\bm Q}\cdot {\bm R}_j}$,
  in addition to a uniform alignment in the $z$  direction, $\la S^z\ra $.
The amplitudes of the spiral, $\la S^+_{A(B)}\ra $, at sub-lattices $A$ and $B$ are given by: 
\begin{eqnarray}
\la S^+_A\ra  &=& \frac{1}{N_c} \sum_{\bm k} 
\la   a_{{\bm k},\ua}^\dagger   a_{\bm{ k+Q},\da}\ra  \nonumber\\
\la S^+_B\ra  &=& \frac{1}{N_c} \sum_{\bm k}
\la   b_{{\bm k},\ua}^\dagger   b_{\bm{ k+Q},\da}\ra .\label{sab}
\end{eqnarray} 
In general, there will be a nonzero angle $\theta$  between the 
transverse sub-lattice  magnetizations 
$\la S^+_A\ra $ and $\la S^+_B\ra $, so that  $\la S^+_B\ra =e^{i\theta}\la S^+_A\ra $.

Following \cite{Auerbach}, the mean field equations are obtained from 
the minimization of the  ground state energy
with respect to the order parameters $\la S^+_{A(B)}\ra $ and $\la S^z\ra $. 
Each  Bloch $\bm k$-state, $\gamma_{\bm k}$, representing an 
elementary excitation,
is  a linear superposition of the fields 
 $  a_{{\bm k},\ua},   a_{{\bm k+\bm Q},\da}, 
   b_{{\bm k},\ua}$ and $  b_{{\bm k+\bm Q},\da}$. Conversely, we
can rewrite each of the fields as a combination  of Bloch states,  and recast
the expectation value  of the kinetic term  in
(\ref{hamilt}) as well as the order parameters (\ref{sab}) 
in terms of $\gamma_{\bm k}$ operators. 
Using Wick's theorem, the expectation value of the Hubbard term
 in the Hamiltonian (\ref{hamilt})
can be expressed  as: 
\begin{eqnarray}
\la H_U\ra =
U \Big( \frac{n^2}{2} - \la S^z_A\ra ^2 - \la S^z_B\ra ^2 -
 \la S^-_A\ra \la S^+_A\ra - \la S^-_B\ra \la S^+_B\ra \Big).\label{medU}
\end{eqnarray} 
By minimizing  $\la H\ra $,  as given in equations
(\ref{hamilt}) and (\ref{medU}), we find that the  Bloch $\bm k$-state, 
$\gamma_{\bm k}$,  diagonalizes  an effective $4\times 4$ 
 Hamiltonian matrix, $H_{eff}(\bm k)$, 
 which can be expressed 
in the basis  $(  a_{{\bm k},\ua},   a_{{\bm k+\bm Q},\da},
   b_{{\bm k},\ua},  b_{{\bm k+\bm Q},\da})$, as
\begin{equation}
H_{eff}(\bm k)=\left[
\begin{array}{cc}
{\cal D}_A(\bm k) & {\cal C}^*(\bm k) \\
{\cal C}(\bm k) &  {\cal D}_B(\bm k)
\end{array} \right]
\end{equation}
where the $2\times 2$ matrices $D_\alpha$ (with $\alpha=A,B$) and $C$ 
are given by:
\begin{equation}
{\cal D}_\alpha(\bm k)=\left[
\begin{array}{cc}
 D({\bm k}) -U\la S^z\ra -\mu &  -U\la S^+_\alpha\ra  \\
 -U\la S^-_\alpha\ra  & D({\bm k+\bm Q}) +U\la S^z\ra -\mu
\end{array} \right]\,, \end{equation}
and
\begin{equation}
{\cal C}(\bm k)=\left[
\begin{array}{cc}
 \phi({\bm k}) & 0  \\
  0 & \phi({\bm k+\bm Q}) 
\end{array} \right].
\end{equation} 
 The mean field equations only determine the phase 
difference between  $\la S^+_A\ra $ and $\la S^+_B\ra $, so we choose  $\la S^+_A\ra $ 
to be real.
At each point in the weak ferromagnetic region of the phase diagram 
we have to  choose the spiral wave vector 
$\bm Q$ that minimizes the ground state energy. This vector should lie along one
of the high symmetry directions in the Brillouin Zone.

We first look for  the most  favorable wave vectors lying along the $\Gamma-X$ 
direction.
For the same parameters as in figure \ref{sw1}, we find a spiral state with 
$\bm Q = \frac{1}{4} \bm{ \Gamma X}$ , $\la S^z\ra = 0.12$ and $\la S^+_A\ra = 0.036$.
We note that this spiraling state  has a smaller $z$-component of
the magnetization  than that in the uniform phase.
The angle between transverse magnetizations $\theta=  0.77\pi\approx 
3\pi/4$.
We find, however, that the energy difference between this spiral state and that
with uniform  magnetization 
is indeed very small, not exceeding $10^{-4}t$ per lattice site.
 This has been 
checked for several lattice sizes.
We also note that the obtained 
spin transverse component  is not negligible compared to  $\la S^z\ra $.
The most favorable spiral wave vector depends slightly on 
interaction and density: 
  at $U=3.5$ and $n=0.8$, for instance,
  $\bm Q = \frac{19}{80}  \bm{ \Gamma X}$ is the most favorable, with
$\la S^z\ra = 0.089$,  $\la S^+_A\ra = 0.046$ and  $\theta=  0.73\pi$. 

\begin{figure}[ht]
\begin{center}
\epsfxsize=8cm
\epsfbox{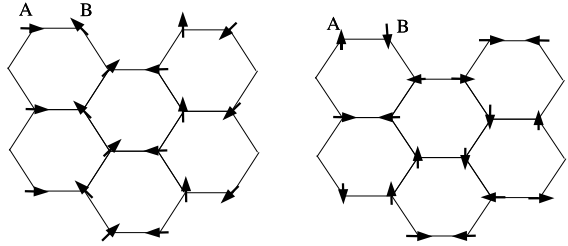}
\caption{ 
Representation of the spin transverse components  $\la S^+_{A,B}\ra $
 for the spiral states with:  $\bm Q= (1/4) \bm{ \Gamma X}$ (left); 
 $\bm Q= (3/4) \bm{ \Gamma K}$ (right). The latter configuration is purely planar
and has the lowest energy
for the same parameters as in figure  \ref{sw1}.}
\label{spir}
\end{center}
\end{figure}

We now consider states with 
$\bm Q \propto \bm{ \Gamma K}$. Overall,  the energies of these spiral states 
are found to be 
 lower than   those of the states with ${\bf Q} \propto {\bf \Gamma X}$ considered above. 
For the same parameters as in figure \ref{sw1}, we find the optimum wave vector
to be ${\bf Q}=\frac{3}{4}{\bf \Gamma K}$, where $\la S^z\ra $ vanishes and
 $\la S^+_A\ra = 0.24= -\la S^+_B\ra $. The spin configuration is, therefore, planar and the 
 two sub-lattices have opposite magnetizations. The energy (per lattice site) 
is $0.02$ lower than that with  uniform magnetization.
An approximate  representation of this state is shown in figure  \ref{spir}.
The length of the most energetically favorable $\bm Q$ depends significantly
 on  $U$ and $n$. If, for instance, $U=3.5$ and  $n=0.8$, then
${\bf Q}=\frac{1}{2}{\bf \Gamma K}$ is the most  favorable, with
$\la S^+_A\ra = 0.18=-\la S^+_B\ra $.  The energy per lattice site of this state
is $0.008$ lower than that of the uniform state. In such a  planar
 spin configuration, a small ferromagnetic 
alignment along $z$ could still arise in the presence of a weak 
anisotropy or external magnetic field.

\begin{figure}[htf]
\begin{center}
\epsfxsize=10cm
\epsfbox{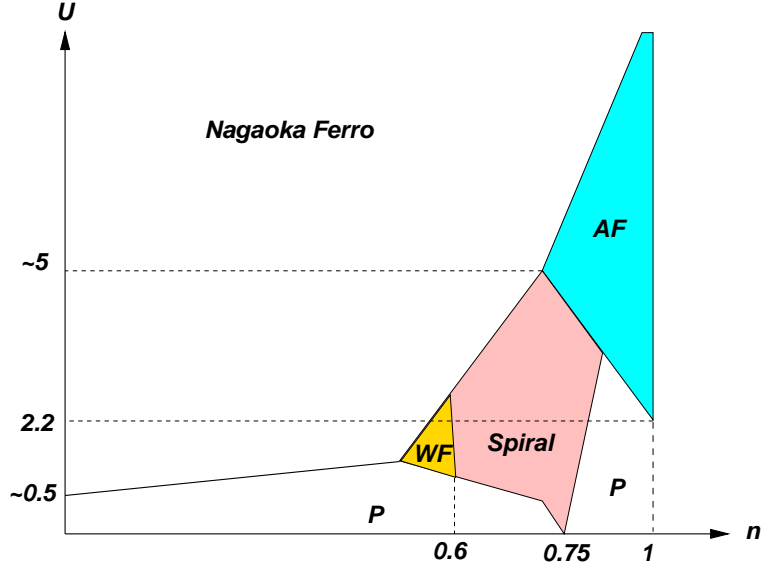}
\end{center}
\caption{Schematic
phase diagram of a  honeycomb layer, with  $t'\ne 0$, showing the 
paramagnetic (P), antiferromagnetic (AF), weak uniform ferromagnetic (WF)
and spiral phases. The Nagaoka phase is a fully polarized ferromagnet. 
The exact position of the transition lines is not given, except
at some special values marked in the axes. The results for 
$t'= 0$ are not qualitatively different.}
\label{fd}
\end{figure}

 An anisotropic perturbation producing an easy axis (or a small magnetic field)
need not  exceed an energy  of  the order $10^{-2}t$ per lattice site
in order to induce the uniform state, specially at the lower $U$
values considered. 
It is still possible that the minimum energy spin structure of the system
could be a superposition of spirals with different $\bm Q$ vectors, as
has been found for the $n=1$ Hubbard triangular lattice
 \cite{fujita}. The search for such structures, as well as their sensitivity to disorder,
is beyond the scope of this work, however.  
We  also find that the spiral states are absent at
 smaller densities ($n < 0.6$). A  schematic representations of our findings
is shown  in figure \ref{fd}.
\vspace{0.5cm}

\section{Disorder as possible mechanism to ferromagnetism}
\label{disorder}

Ferromagnetic order induced by disorder in proton irradiated 
graphite was observed by Y. Kopelevich {\it et al.} \cite{kopelevich1}.
The disorder induced by proton irradiation can be modeled by diagonal
disorder, where the local energy of some sites is modified.
The problem of treating disorder and
Coulomb interaction together is a hard one in condensed matter physics
\cite{ziegler}.
Therefore we start by studying the effect of a finite density of uncorrelated
 impurities on the
electron gas in the honeycomb lattice at half filling, where 
the low-energy electronic excitations can be described 
by massless Dirac fermions
\cite{guinea}. The effect of disorder on the properties of 
Dirac fermions leads to some unexpected results, and was discussed
in the context of disordered  superconductors 
by some authors \cite{lee,araujo}.
In our study, we compute the $T-$matrix using the massless Dirac fermion
description.

The Matsubara
Green's functions are determined via the equation-of-motion method.
After the usual manipulations \cite{doniach} we obtain the $2\times 2$
Green's function matrix as 
\begin{equation}
\bm G(\bm p,i\omega_n)=\bm G^0(\bm p,i\omega_n)+\bm G^0(\bm p,i\omega_n)
\bm T(i\omega_n)\bm G^0(\bm p,i\omega_n)\,,
\end{equation}
with the Matsubara $T-$matrix  given by
\begin{equation}
\bm T(i\omega_n)=\frac V N [\bm 1-\frac{V}{N} \bar {\bm G}^0(i\omega_n)]^{-1}\,,
\label{1imp}
\end{equation}
where $\bm G^0(\bm p,i\omega_n)$ is the Green's function matrix
with $V=0$, and 
\begin{equation}
\bar {\bm G}^0(i\omega_n)=\sum_{\bm p}\bm G^0(\bm p,i\omega_n)\,.
\end{equation}
For a small density, $n_{imp}$, of scatterers (but finite in the
thermodynamic limit), and after the position of the scatterers
has been averaged over ensemble configurations,
 the Green's function matrix can be written as
\begin{equation}
\bm G(\bm p,i\omega_n)=
[[\bm G^0(\bm p,i\omega_n)]^{-1}-\bm \Sigma(i\omega_n)]^{-1}\,,
\label{gtm}
\end{equation}
where \cite{doniach}
\begin{equation}
\bm \Sigma(i\omega_n)= V n_{imp} 
[\bm 1-\frac{V}{N} \bar {\bm G}^0(i\omega_n)]^{-1}\,.
\end{equation}
Once the Green's function (\ref{gtm}) is known, one can proceed
to include the effect of correlations into the problem. The
electronic bound states are given by the poles of  (\ref{1imp}).
There always is  a bound state due to the impurity, below the energy band, 
independently of the value of $V$. 

\begin{figure}[ht]
\begin{center}
\includegraphics*[width=8cm]{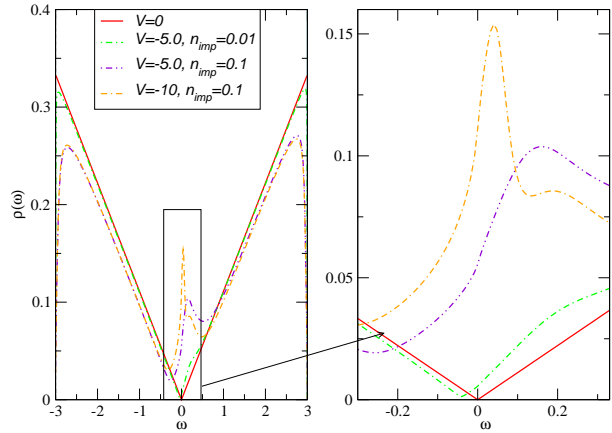}
\caption{
Density of states $\rho(\omega)$ for several impurity
densities $n_{imp}$ and $V$ values, in a model of Dirac fermions.
The left panel show $\rho$ over the  bandwidth. 
The  right panel
shows a zoom of the density of states presented in the
left panel.}
\label{dosV}
\end{center}
\end{figure}

The existence of bound states allows for a possible mechanism to the disorder
induced ferromagnetic behaviour in proton irradiated graphite. 
Graphite is usually modeled as 
a half-filled honeycomb plane, where electrons near the Fermi level have linear
dispersion \cite{guinea}.  
The sample irradiation produces the displacement of the carbon
atoms from their original position. In this case,
even if  hydrogen atoms become bonded to
some of the carbons, from the lattice point of view
a dilution of lattice points is being induced. In this case, we should
take the limit $V\rightarrow\infty$, even if $n_{imp}$ is small. This
effect leads to drastic change in the density of states $\rho(\omega)$, 
where a strong enhancement of $\rho(\omega)$ in the vicinity of $\omega=0$
is obtained. Such an enhancement can be responsible for a large
reduction of the critical $U$ needed for ferromagnetism, 
as follows from the  Stoner criterion.  
The effect of disorder on the density of states of  Dirac fermions 
is shown  in figure \ref{dosV} for several values
of $n_{imp}$ and $V$. If $V$ is negative there are bound states
for the electrons below the bottom of the band. As $V$ increases
an enhancement of $\rho(\omega)$ in the vicinity of $\omega=0$
starts to develop. 
We have, therefore, two possible routes toward the appearance of magnetic
order in graphite, namely bound states and enhancement of the density
of states. This type of enhancement
of  $\rho(\omega)$ is characteristic of acoustic excitations, either
fermionic (Dirac fermions) or bosonic (magnons)\cite{chernyshev}.
This very qualitative view has to be corroborated by more detailed 
calculations taking into account both disorder and Coulomb interaction. 
In  particular, it is important to compute how the critical lines from the
paramagnetic  to the magnetic phases change with the amount of disorder.
These issues will be  the subject of a future publication.


\section{Conclusion}

In conclusion, we have studied the spin  collective excitations 
of the  homogeneous weak ferromagnetic state in the honeycomb lattice and
found an instability to spiral spin structures at electron densities above
$n\approx 0.6$  
Although our calculation is performed for a system with 
the same type of atoms it is simple to generalize it for 
honeycomb lattices with different type of atoms as
in BNC hexagonal sheets \cite{susumu}.
However, the main differences will be: (i) the number of
optical and acoustic branches increases; (ii) the 
different site energies due to different  atoms
will change the form of the spin wave bands.
We have also suggested a possible mechanism for ferromagnetism  
in irradiated graphite: the appearance of bound states due to disorder 
and the  enhancement of the density  of states. 
If it becomes possible in the future to perform
neutron scattering experiments on ferromagnetic graphite allotrope
the results we present here may have direct experimental importance.


\section*{Acknowledgments}
The authors would like to acknowledge A. H. Castro Neto
for suggesting the $T-$matrix calculation and for many useful
discussions on disordered systems.
N.M.R.P is thankful to the Quantum Condensed Matter
Visitor's Program at Boston University and
Funda\c{c}\~ao para a Ci\^encia e Tecnologia for a sabbatical grant
partially supporting his sabbatical leave.



\section*{References}
\begin{thebibliography}{100}

\bibitem{anderson}
P. W. Anderson, Science {\bf 235}, 1196 (1987)

\bibitem{kopelevich1} 
Y. Kopelevich, J. H. S. Torres, R. R. da Silva, F.  Mrowka, H. Kempa
and P. Esquinazi,      Phys. Rev. Lett. {\bf 90}, 156402 (2003)

\bibitem{matthew}
L. Balents, M. P. A.  Fisher and S. M. Girvin,  Phys. Rev. B
{\bf 65}, 224412 (2002)

\bibitem{esquinazi} 
P. Esquinazi, A. Setzer, R. H\"ohne, C. Semmelhack, Y. Kopelevich, 
D. Spemann, T.  Butz, B. Kohlstrunk  and  M. L\"osche, 
   Phys. Rev.  B {\bf 66}, 24429 (2002);
P. Esquinazi, D. Spemann, R.   H\"ohne, A. Setzer, K. H. Han and T. Butz, 
    Phys. Rev. Lett. {\bf 91}, 227201 (2003).


\bibitem{tanaka}
K. Takada, H.  Dakurai, E. Takayama-Muromachi,
F. Izumi, R. A.  Dilinian and  T. Sasaki
   Nature {\bf 422}, 53 (2003)

\bibitem{kopelevich2}
 Y. Kopelevich, P. Esquinazi, J. H. S. Torres and S. Moehlecke,
J. Low Temp. Phys. {\bf 119}, 691 (2000).

\bibitem{makarova}
T. Makarova, B.  Sundqvist, R.  H\"ohne, 
P.  Esquinazi,Y.  Kopelevich, P.  Scharff,
V. A. Davydov, L. S. Kashevarova and A. V. Rakhmanina,
   Nature {\bf 413}, 716 (2001)

\bibitem{he}
T. He and R. J. Cava, Phys. Rev. B {\bf 63}, 172403 (2001).


\bibitem{novoselov}
K. S. Novoselov, A. K. Geim, S. V. Morozov,
D. Jiang, Y. Zhang, S.V. Dubonos, I. V. Grigorieva and A. A. Firsov,
Science {\bf 306}, 666 (2004).

\bibitem{zhang}
Y. Zhang, J. P. Small, W. V. Pontius and P. Kim
cond-mat/410314.

\bibitem{berger}
C. Berger, Z. Song, T. Li, X. Li, A. Y. Ogbazghi and R. Feng,
cond-mat/410240.

\bibitem{mattis}
D. C. Mattis,    Phys. Rev. {\bf 132}. 2521 (1964)

\bibitem{yamada}
 H. Yamada  and M. Shimizu,   J. Phys. Soc. Japan {\bf 22}, 1404
(1967)


\bibitem{prb04}
N. M. R. Peres, M. A. N. Ara\'ujo and Daniel Bozi,
    Phys. Rev. B {\bf 70},
195122 (2004)

\bibitem{guinea}
J. Gonzalez, F. Guinea and M. A. H. Vozmediano, 
Nucl. Phys. B {\bf 424},  595 (1994)
J. Gonzalez, F. Guinea, and M. A. H. Vozmediano, Phys. Rev. Lett.
{\bf 77}, 3589 (1996).

\bibitem{tesa} 
A. Singh  and  Z. Tesanovic,   Phys. Rev. B {\bf 41},
11457 (1990);   {\bf 41}, 614 (1990)

\bibitem{Auerbach} 
A. Auerbach,  {\it Interacting Electrons and Quantum Magnetism}, (New York,
Springer-Verlag, 1994) p.40.

\bibitem{fujita} M. Fujita, T. Nakanishi and K. Machida,
  Phys. Rev. B  {\bf 45}, 2190 (1992)

\bibitem{ziegler}
W. Ziegler, D. Poilblanc, R. Preys, W. Hanke and D. J. Scalapino,
Phys. Rev. B {\bf 53}, 8704 (1996).

\bibitem{lee}
Patrick A. Lee,  Phys. Rev. Lett. {\bf 71}, 1887 (1993).

\bibitem{araujo}
M.A.N. Ara\'ujo,	
Int. J. of Mod. Phys. B {\bf 15},  409 (2001). 

\bibitem{doniach}
S. Doniach and E. H. Sondheimer, {\it Green's functions for solid
state physicists}, (London, Imperial College Press, 1988) chaps. 4 and 5.


\bibitem{chernyshev}
A. L. Chernyshev, Y. C. Chen and A. H. Castro Neto, 
Phys. Rev. Lett. {\bf 87}, 067209 (2001); 
{\it idem}
 Phys. Rev. B {\bf 65}, 104407 (2002).

\bibitem{susumu}
S.  Okada  and A. Oshiyama,   Phys. Rev. Lett. {\bf 87}, 146803 (2001)

\end{thebibliography}
\end{document}